\documentstyle[12pt]{article}
\newcommand{\prd}[3]{{\it Phys. Rev.} {\bf D#1}, #2 (19#3)}
\newcommand{\pl}[3]{{\it Phys. Lett.} {\bf #1B},#2  (19#3)}
\newcommand{\np}[3]{{\it Nucl. Phys.} {\bf B#1},#2  (19#3)}
\newcommand{\prl}[3]{{\it Phys. Rev. Lett.} {\bf #1}, #2  (19#3)}

\newcommand{\ra}{\rightarrow}

\newcommand{\matel}[3]{\left<#1\right|#2\left|#3\right>}
\newcommand{\cp}{{\cal CP}}

\newcommand{\cpv}{\not\!\!{\cal CP}}
\newcommand{\T}{{\cal T}}

\relax
\begin{document}
\begin{titlepage}
\def\ba{\begin{array}}
\def\ea{\end{array}}
\def\thefootnote{\fnsymbol{footnote}}
\vfill
\hskip 4in AMES-HET-94-13

\hskip 4in November, 1994
\vspace{1 in}
\begin{center}
{\large \bf $\cp$ Violation in Hyperon Decays}\footnote{Invited talk
presented at HQ94, Oct.7-10, 1994, Charlottesville, Virginia.}
\\

\vspace{1 in}
{\bf  G.~Valencia}\\
{\it           Department of Physics,
               Iowa State University,
               Ames IA 50011}\\
\vspace{1 in}
     %	{\large \bf ABSTRACT}
\end{center}

\begin{abstract}
In this talk we review the status of the theoretical estimates for
$\cp$ violating asymmetries in non-leptonic hyperon decays.
\end{abstract}

\end{titlepage}

\section{-- The Decay $\Lambda^0 \ra p \pi^-$}

The reaction $\Lambda^0 \ra p \pi^-$ will be used as an example to set up
the model independent formulation of the $\cp$ odd observables in
non-leptonic hyperon decays of the form $B_i \ra B_f \pi$.
In the $\Lambda^0$ rest frame, $\vec{\omega}_{i,f}$ will denote unit vectors in
the directions of the $\Lambda$ and $p$ polarizations, and
$\vec{q}$ the proton momentum.
The isospin of the final state is
$I= 1/2 {\rm ~or~} 3/2$, and each of these two states can be reached
via a $\Delta I = 1/2 {\rm ~or~} 3/2$ weak transition respectively.
There are also two possibilities for the parity of the final state.  They are
the $s$-wave, $l=0$, parity odd state (thus reached via a parity
violating amplitude); and the $p$-wave, $l=1$, parity even state reached
via a parity conserving amplitude.

We first perform a model independent analysis of the decay by writing the
most general matrix element consistent with Lorentz
invariance: \cite{tdlee,marshak,commins}
\begin{equation}
{\cal M} = G_F m_\pi^2 \overline{U}_P(A-B\gamma_5)U_\Lambda .
\label{gmatel}
\end{equation}
This matrix element reduces to
\begin{equation}
A(B_i \ra B_f \pi) = s + p \vec{\sigma}\cdot\vec{q}
\label{nonrel}
\end{equation}

In terms of these quantities one can compute the decay distribution, and the
total decay rate. One finds that the decay is characterized by three
independent observables: the total decay rate and two parameters that
determine the angular distribution.
The total decay rate is given by:
\begin{equation}
\Gamma = {|\vec{q}|(E_P+M_P) \over 4 \pi M_\Lambda}G_F^2 m_\pi^4
\left(|s|^2+|p|^2\right).
\label{drate}
\end{equation}
The angular distribution is proportional to:
\begin{equation}
{d\Gamma \over d\Omega} \sim 1+
\gamma\vec{\omega}_i\cdot\vec{\omega}_f + (1-\gamma)\hat{q}\cdot\vec{\omega}_i
\hat{q}\cdot\vec{\omega}_f
+ \alpha \hat{q}\cdot(\vec{\omega}_i+\vec{\omega}_f)
+ \beta \hat{q}\cdot(\vec{\omega}_f\times\vec{\omega}_i),
\label{msqsim}
\end{equation}
where we have introduced the notation:
\begin{equation}
\alpha  \equiv {2 {\rm Re} s^*p \over |s|^2 + |p|^2},  \;\;
\beta  \equiv {2 {\rm Im} s^*p \over |s|^2 + |p|^2}, \;\;
\gamma  \equiv {|s|^2 - |p|^2 \over |s|^2 + |p|^2} .
\label{albega}
\end{equation}
However, only two of these parameters are independent, since $\alpha^2
+\beta^2 + \gamma^2=1$. We will treat $\alpha$ and $\beta$ as the
independent ones, although sometimes the parameters
$\alpha$ and $\phi$, with $\beta=\sqrt{1-\alpha^2}\sin\phi$ and
$\gamma=\sqrt{1-\alpha^2}\cos\phi$ are used instead.

The parameter $\alpha$ governs the $T$-even correlation between the proton
momentum and the $\Lambda$ polarization, whereas $\beta$ governs the $T$-odd
correlation involving the two polarization vectors and the proton momentum.
I use $T$ to indicate
the operation that reverses the sign of all momenta and spins in the
reaction, and {\bf not} the time reversal operation. The significance
of this discrete symmetry is that operators that are even under it
can only be used to construct
$\cp$ odd observables that require final state interactions,
whereas those that are odd can be used to construct $\cp$ odd observables
that do not vanish in the absence of final state interactions.

One way to interpret the parameter $\alpha$ follows from considering the
angular distribution in the case when the final baryon polarization
is not observed:
\begin{equation}
{d\Gamma \over d\Omega}={1 \over 16 \pi^2}{|\vec{q}|\over M_\Lambda}
G_F^2m_\pi^4(E_P+M_P)\left(|s|^2+|p|^2\right)
\left(1+\alpha \hat{q}\cdot \vec{\omega}_i\right).
\label{alonly}
\end{equation}
The polarization of the decay proton in the $\Lambda^0$ rest frame is given by:
\begin{equation}
\vec{\cal P}_P = {1 \over 1+\alpha\vec{\cal P}_\Lambda\cdot\hat{q}}
\biggl[\left(\alpha+\vec{\cal P}_\Lambda\cdot\hat{q}\right)\hat{q}
+\beta
\left(\vec{\cal P}_\Lambda\times\hat{q}\right)+\gamma\left(\hat{q}
\times\left(\vec{\cal P}_\Lambda\times\hat{q}\right)\right)\biggr].
\label{polp}
\end{equation}
{}From this expression we can relate $\beta$ to the proton polarization
in the direction perpendicular to the plane formed by
the $\Lambda$ polarization and
the proton momentum.
If the initial hyperon is unpolarized,
$\alpha$ gives us the polarization of the proton:
\begin{equation}
\vec{\cal P}_p = \alpha_\Lambda \hat{q}
\label{otheral}
\end{equation}

Since the proton polarization is not measured, the parameter $\beta$
is not useful for the reaction $\Lambda \ra p \pi^-$. It is, however,
useful for other hyperon decays, such as the chain:
\begin{equation}
\Xi^- \ra \Lambda^0 \pi^- \ra p \pi^- \pi^-
\label{chain}
\end{equation}
where the second decay analyzes the polarization of the $\Lambda$
and allows one to observe the parameter $\beta_{\Xi}$.

\section{-- $\cp$-odd Observables}

To construct $\cp$-odd observables we compare the reactions
$\Lambda^0 \ra p \pi^-$ and $\overline{\Lambda}^0 \ra \overline{p} \pi^+$
in terms of the three independent observables. One can show that
$\cp$ symmetry predicts that:
\begin{eqnarray}
\overline{\Gamma} &=& \Gamma \nonumber \\
\overline{\alpha} &=& - \alpha \nonumber \\
\overline{\beta} &=& - \beta
\label{cppred}
\end{eqnarray}
and, therefore, one can construct the following $\cp$-odd observables:
\cite{donpa}
\begin{eqnarray}
\Delta &\equiv & {\Gamma - \overline\Gamma \over \Gamma + \overline\Gamma}
\nonumber \\
A & \equiv &{\alpha \Gamma + \overline{\alpha}\overline{\Gamma} \over
\alpha \Gamma - \overline{\alpha}\overline{\Gamma}}
\approx {\alpha + \overline{\alpha} \over \alpha - \overline{\alpha}}
+ \Delta \nonumber \\
B &\equiv &{\beta\Gamma + \overline{\beta}\overline{\Gamma} \over
\beta\Gamma - \overline{\beta}\overline{\Gamma}}
\approx {\beta +\overline{\beta} \over \beta - \overline{\beta}}+\Delta
\label{cpobs}
\end{eqnarray}

One can show that in the low energy reaction $p\overline{p} \ra
\Lambda \overline{\Lambda} \ra p_f \pi^- \overline{p}_f \pi^+$ it
is possible to construct counting asymmetries that measure $A$ and $B$.
\cite{dhv} If we  label each particle's momentum by the particle
name, and denote by $N^\pm_p$ the number of events with
$(\vec{p}_i\times\vec{p}_\Lambda)\cdot\vec{p}_f$ greater or less than
zero then: \cite{dhv}
\begin{equation}
\overline{A} = {N_p^+ -N^-_P + N_{\overline{p}}^+ -N_{\overline{p}}^- \over
N_{total} } = {\cal P}_\Lambda \alpha_\Lambda A_\Lambda
\label{countal}
\end{equation}
Similarly, in the reaction $p \overline{p} \ra \Xi \overline{\Xi} \ra
\Lambda \pi^- \overline{\Lambda}\pi^+ \ra p_f \pi^-\pi^-\overline{p}_f
\pi^+\pi^+$, we can define $\overline{N}^\pm_p$ to be the number
of events with ${\cal P}_{\Xi}\cdot(\vec{p}_f\times\vec{p}_\Lambda)$
greater or less than zero and construct the counting asymmetry: \cite{dhv}
\begin{equation}
\overline{B} = {\overline{N}_p^+ -\overline{N}^-_P +
\overline{N}_{\overline{p}}^+ -
\overline{N}_{\overline{p}}^- \over
N_{total} } = {\pi \over 8}{\cal P}_\Xi \alpha_\Lambda
\beta_\Xi(A_\Lambda + B_\Xi)
\label{countbe}
\end{equation}

\section{-- Isospin Decomposition}

To discuss the final state interaction phases it is
convenient to analyze the
final pion-nucleon system in terms of isospin and parity eigenstates. In
that way we can have in mind the simple picture:
\begin{equation}
\Lambda^0 \begin{array}{c} t=0 \\ \longrightarrow \\H_w \end{array}
\bigg(p\pi\bigg)^I_\ell
\begin{array}{c} H_s \\ \longrightarrow \\ \delta^I_\ell \end{array}
\bigg(p\pi\bigg)^I_\ell .
\label{naive}
\end{equation}
At $t=0$ the weak Hamiltonian induces the decay of the $\Lambda^0$
into a pion-nucleon system with isospin and parity given by $I,~\ell$.
If there is $\cp$ violation in this decay there will be a $\cp$-odd
phase $\phi^I_\ell$.
This pion-nucleon system is
an eigenstate of the strong interaction.
Furthermore, at an energy equal to the
$\Lambda$ mass, it is the only state with these quantum numbers. The
pion-nucleon system will then rescatter due to the strong interactions into
itself, and in the process pick up a phase $\delta^I_\ell$. This is
an example of what is known as Watson's theorem.

If we parameterize the amplitudes in non-leptonic hyperon decays as:
\begin{equation}
s =\sum_I s_I e^{i(\delta^I_s+\phi^I_s)}\;\;
p =\sum_I p_I e^{i(\delta^I_p+\phi^I_p)},
\label{newpar}
\end{equation}
then $\cp\T$ invariance of the weak Hamiltonian predicts:
\begin{equation}
\overline{s} =\sum_I -s_I e^{i(\delta^I_s-\phi^I_s)}\;\;
\overline{p} =\sum_I p_I e^{i(\delta^I_p-\phi^I_p)},
\label{cpthyp}
\end{equation}
whereas $\cp$ invariance of the weak interactions predicts:
\begin{equation}
\overline{s} =\sum_I -s_I e^{i(\delta^I_s+\phi^I_s)}\;\;
\overline{p} =\sum_I p_I e^{i(\delta^I_p+\phi^I_p)}.
\label{cphyp}
\end{equation}
{}From this we see that the $\phi^I_\ell$ phases violate $\cp$.
We want to extract $\phi^I_\ell$ from the $\cp$-odd observables
discussed in the previous section.

Introducing the notation used in the literature: \cite{overpa}
\begin{eqnarray}
S(\Lambda^0_-)= - \sqrt{2/3}S_{11}e^{i(\delta_1+\phi^s_1)}+\sqrt{1/3}
S_{33}e^{i(\delta_3+\phi^s_3)}\nonumber \\
P(\Lambda^0_-)= - \sqrt{2/3}P_{11}e^{i(\delta_{11}+\phi^p_1)}+\sqrt{1/3}
P_{33}e^{i(\delta_{33}+\phi^p_3)}\nonumber \\
S(\Xi^-_-)= S_{12}e^{i(\delta_2+\phi^s_{12})}+{1 \over 2}S_{32}
e^{i(\delta_2+\phi^s_{32})}\nonumber \\
P(\Xi^-_-)= P_{12}e^{i(\delta_{21}+\phi^p_{12})}+{1 \over 2}P_{32}
e^{i(\delta_{21}+\phi^p_{32})}
\label{overnot}
\end{eqnarray}
where $\Lambda^0_-$ refers to the reaction $\Lambda^0 \ra p \pi^-$ and
$\Xi^-_-$ refers to the reaction $\Xi^- \ra \Lambda^0 \pi^-$.
The notation for the isospin amplitudes is $S_{ij} \equiv
S_{2\Delta I,2I}$, $P_{ij} \equiv P_{2\Delta I,2I}$,
the $s$-wave phases are denoted by $\delta_{2I}$ and the $p$-wave
phases by $\delta_{2I,1}$.

It is useful to construct approximate expressions based
on the fact that there are three small parameters in the problem:
\begin{itemize}
\item The strong rescattering phases are measured or estimated to be small.
Experimentally we know that \cite{roper}
\begin{equation}
\delta_1 \approx 6.0^\circ,\;\;\delta_3 \approx -3.8^\circ,\;\;
\delta_{11}\approx -1.1^\circ,\;\;\delta_{31}\approx-0.7^\circ
\label{stphex}
\end{equation}
with all the errors on the order of $1^\circ$. For the $\Xi$ decays
there are no experimental results. An early calculation within
a model predicted
$\delta_{21}=-2.7^\circ$ and $\delta_2 = -18.7^\circ$. \cite{nath}
A recent
calculation using chiral perturbation theory predicts instead
$\delta_{21}=-1.7^\circ $ and $\delta_2=0$. \cite{lusawi}
Clearly the resulting
asymmetries will be completely different depending on which of these
results is closer to the true scattering phases.

\item The $\Delta I=3/2$ amplitudes are much smaller than the
$\Delta I=1/2$ amplitudes. Experimentally we know that: \cite{overseth}
\begin{eqnarray}
S_{33}/S_{11} = 0.027 \pm 0.008, \;\; P_{33}/P_{11}=0.03\pm 0.037
\nonumber\\
S_{32}/S_{12} = -0.046 \pm 0.014,\;\;P_{32}/P_{12}=-0.01\pm0.04
\label{expamp}
\end{eqnarray}

\item The $\cp$ violating phases are presumed to be small.
\end{itemize}

To leading order in all the small quantities one finds: \cite{donpa}
\begin{eqnarray}
\Delta (\Lambda^0_-) &=& \sqrt{2}{S_{33}\over S_{11}}
\sin\left(\delta_3-\delta_1\right)
\sin\left(\phi^s_3-\phi^s_1\right)\nonumber \\
A(\Lambda^0_-)&=& -\tan\left(\delta_{11}-\delta_1\right)
\sin\left(\phi^p_1-\phi^s_1\right)\nonumber \\
B(\Lambda^0_-)&=& \cot\left(\delta_{11}-\delta_1\right)
\sin\left(\phi^p_1-\phi^s_1\right)
\label{approxas}
\end{eqnarray}
and
\begin{eqnarray}
\Delta (\Xi^-_-) &=& 0 \nonumber \\
A(\Xi^-_-)&=& -\tan\left(\delta_{21}-\delta_2\right)
\sin\left(\phi^p_{12}-\phi^s_{12}\right)\nonumber \\
B(\Xi^-_-)&=& \cot\left(\delta_{21}-\delta_2\right)
\sin\left(\phi^p_{12}-\phi^s_{12}\right)
\label{approxasxi}
\end{eqnarray}

We can see in these expressions that $\Delta$ arises mainly from an
interference between a $\Delta I=1/2$ and a $\Delta I=3/2$ $s$-waves,
and that it is suppressed by three small quantities. On the other
hand, $A$ arises as an interference of $s$ and $p$-waves of the same
isospin  and, therefore, it is not suppressed by
the $\Delta I =1/2$ rule. Finally, we can see that $B$ is not suppressed
by the small rescattering phases. This is as we expected for a $\cp$ odd
observable that is also (naive)-$T$ odd. The hierarchy $B >> A >> \Delta$
emerges. \cite{donpa} The quantity $\Delta(\Xi^-_-)$ vanishes because
there is only one isospin final state in this decay.

This is as far as we can go in a model independent manner. If we want to
predict the value of these observables within a model for $\cp$ violation
we take the value for the ratio of amplitudes  and for
the strong rescattering phases
from experiment and we try to compute the weak phases from theory.

\section{-- Standard model calculation}

In the case of the minimal standard model, the $\cp$ violating phase resides
in the CKM matrix. For low energy transitions, this phase shows up as the
imaginary part of the Wilson coefficients in the effective weak
Hamiltonian. In the notation of Buras \cite{buras},
\begin{equation}
H_W^{eff} = {G_F \over \sqrt{2}}V^*_{ud}V_{us}\sum_i c_i(\mu)Q_i(\mu) +
{\rm ~hermitian~conjugate}
\label{effweak}
\end{equation}
$Q_i(\mu)$ are four quark operators, and $c_i(\mu)$ are the Wilson
coefficients that are usually written as:
\begin{eqnarray}
c_i(\mu) &=&z_i(\mu)+\tau y_i(\mu) \nonumber \\
\tau &=& - {V^*_{td}V_{ts} \over V^*_{ud}V_{us}}
\end{eqnarray}
with the $\cp$ violating phase being the phase of $\tau$.
Numerical values for these coefficients can be found, for
example, in Buchalla {\it et. al.} \cite{buchalla}

The calculation would proceed as usual, by evaluating the hadronic matrix
elements of the four-quark operators in Eq.~\ref{effweak} to obtain real
and imaginary parts for the amplitudes, schematically:
\begin{equation}
\matel{p \pi}{H_w^{eff}}{\Lambda^0}|^I_\ell = {\rm Re}M^I_\ell + i {\rm Im}
M^I_\ell ,
\label{schematic}
\end{equation}
and to the extent that the $\cp$ violating phases are small, they can be
approximated by
\begin{equation}
\phi^I_\ell \approx { {\rm Im}M^I_\ell \over {\rm Re}M^I_\ell}.
\label{smallph}
\end{equation}
At present, however, we do not know how to compute the matrix elements so
we cannot actually implement this calculation. If we try to follow what is done
for kaon decays, we would compute the matrix elements using factorization
and vacuum saturation as a reference point, then define some parameters
analogous to $B_K$ that would measure the deviation of the matrix elements
from their vacuum saturation value. A reliable calculation of the ``$B$''
parameters would probably have to come from lattice QCD.

For a simple estimate, we can take the real part of the matrix
elements from experiment (assuming that the measured amplitudes are real,
that is, that $\cp$ violation is small), and compute the imaginary
parts in vacuum saturation. This approach provides a conservative
estimate for the weak phases because the model calculation of the
real part of the amplitudes is much smaller than the experimental value.
Of course, if we cannot predict the real part of the amplitude
at all, we might question the reliability of the imaginary part as
well.

There are many models in the literature that
claim to fit the experimentally measured amplitudes. Without entering into
the details of these models, it is obvious that to fit the data,
the models must enhance
some or all of the matrix elements with respect to vacuum saturation. Clearly,
one would get completely different phases depending on which matrix elements
are enhanced. It is not surprising, therefore, that a survey of these models
yields weak $\cpv$ phases that differ by an order of magnitude \cite{steger}.

The approximate weak phases estimated in vacuum saturation are: \cite{steger}
\begin{eqnarray}
\phi^1_s &\approx& -3 y_6 {\rm Im}\tau \nonumber \\
\phi^1_p &\approx& -0.3 y_6 {\rm Im}\tau \nonumber \\
\phi^3_s &\approx& \left[3.56(y_1+y_2)+4.1(y_7+2y_8)
{m_\pi^2 \over m_s (m_u+m_d)}\right] {\rm Im}\tau
\label{approxph}
\end{eqnarray}
To get some numerical estimates we use the values for the
Wilson coefficients of Buchalla {\it et. al.} \cite{buchalla} with
$\mu=1$~GeV, $\Lambda_{QCD}=200$~MeV. Although quantities such as the
quark masses that appear in Eq.~\ref{approxph} are not physical \cite{dghb},
we will use for an estimate the value $m_\pi^2/(m_s(m_u+m_d))\sim 10$.
For the quantity ${\rm Im}\tau$ we use the current upper bound
${\rm Im}\tau \leq 0.0014$. Putting all the numbers together yields:
\begin{eqnarray}
\Delta (\Lambda^0_-) &=& \cases{-1.4 \times 10^{-6} & for $m_t=150$~GeV \cr
                  -9.1 \times 10^{-7} & for $m_t=200$~GeV \cr}\nonumber \\
A(\Lambda^0_-) &=& 3.7 \times 10^{-5} \nonumber \\
B(\Lambda^0_-) &=& 2.4 \times 10^{-3}
\label{vsnumres}
\end{eqnarray}
A poor man approach to the problem of the hadronic matrix elements
consists of surveying several models. Combining this
with a careful analysis of the allowed range for the short distance
parameters that enter the calculation yields results similar
to that of Eq.~\ref{vsnumres}: that
$A$ is in the range of ${\rm ``a~few"~}\times 10^{-5}$ and that
$\Delta$ is almost two orders of magnitude smaller. The rate asymmetry
exhibits a strong dependence on the top-quark mass: for a certain
value of $m_t$,
the two terms in Eq.~\ref{approxph} cancel
against each other. The angular
correlation asymmetries, on the other hand, depend mildly on the top-quark
mass. This is understood from the point of view that the most important
effect of a large top-quark mass is to enhance electroweak corrections
to the effective weak Hamiltonian. This is important for the $\Delta I=
3/2$ amplitudes but not for the $\Delta I =1/2$ amplitudes.

\section{-- Other Models of $\cp$ Violation}

Other models of $\cp$ violation contain additional short distance operators
with $\cp$ violating phases. \cite{wein,moha}
Some of these have been analyzed in the literature
fixing the strength of the $\cp$ violating couplings from the parameter
$\epsilon$ in kaon decays. \cite{donpa} A summary of those results is shown in
Table~1, taken from a recent talk by He. \cite{hedpf}

Table 1. Sample of models of CP violation in hyperon decays.

\begin{tabular}{|c|c|c|c|}
\hline
$\Lambda\;\mbox{decay}$&KM model&Weinberg Model& Left-Right Model\\ \hline
$\Delta(\Lambda^0_-)$& $<10^{-6}$&$-0.8\times 10^{-5}$&0\\
$A(\Lambda^0_-)$& $-(1\sim 5)\times 10^{-5}$& $-2.5\times 10^{-5}$&$-1.1\times
10^{-5}$\\
$B(\Lambda^0_-)$& $(0.6\sim 3)\times 10^{-4}$&
$1.6\times 10^{-3}$&$7.0\times 10^{-4}$\\ \hline
$\Xi\; \mbox{decay}$&&&\\ \hline
$\Delta(\Xi^-_-)$& 0&0&0\\
$A(\Xi^-_-)$&$-(1\sim 10)\times 10^{-5}$&$-3.2\times 10^{-4}$&$2.5\times
10^{-5}$\\
$B(\Xi^-_-)$&$(1\sim 10)\times 10^{-3}$&$3.8\times 10^{-3}$&$-3.1\times
10^{-4}$\\ \hline
\end{tabular}
\vspace{5mm}

An important point is that the numbers for the $\Xi^- \ra \Lambda \pi^-$
decay were obtained using the early model calculation. \cite{nath} With the
chiral perturbation theory numbers \cite{lusawi} $A(\Xi^-_-)$ would be smaller
by a factor of 10.

\section*{Acknowledgements}
I would like to thank J. F. Donoghue, X. G. He, B. R. Holstein and H. Steger
for pleasant collaborations on the subject of this talk. This work was
supported in part by a DOE OJI award.

\end{document}